# End effects in optical fibres

Jeremy Allington-Smith[1], Colin Dunlop, Ulrike Lemke[2]

and Graham Murray

Centre for Advanced Instrumentation, Durham University, Physics Dept, Rochester Building, South Rd, Durham DH1 3LE, UK

## Abstract

The performance of highly-multiplexed spectrographs is limited by focal degradation (FRD) in the optical fibres. It has already been shown that this is caused mainly by processes concentrated around the mounting points at the ends of the fibres. We use the thickness of rings produced in the farfield when a fibre is illuminated by a collimated beam, to estimate the size of the region where the FRD is generated. This requires the development of a new model, using features of existing ray-tracing and wave-based models, which fits existing data very well. The results suggest that the amount of FRD is primarily determined by the length of fibre bonded into the supporting ferrule. We point out the implications for the production of future fibre systems.

---

[1] j.r.allington-smith@durham.ac.uk

[2] Now at the University of Göttingen

# 1 Introduction

Optical fibres are key technology for highly-multiplexed and precision spectroscopy. Despite their imperfections as optical components (internal absorption and non-conservation of Etendue) and the difficulty of manufacturing fibre systems, they remain the only practical option for multiplexed spectroscopy over large fields of view (e.g. 2dF - Colless et al. 2001; FMOS – Kimura et al. 2010). They continue to be important to Extremely Large Telescopes and to 4m-class telescopes for widefield spectroscopy to address specific cosmological issues (e.g. DESI, Schlegel et al. 2011). These require spectral resolving power $R \approx 4000$ and multiplex gain, $G \approx 5,000$ while "Galactic Archaeology" applications require $R \approx 20,000$ and exoplanetary studies require $R > 100,000$ (e.g. Wilken et al. 2010). Here the dominant issue is the scrambling properties of the fibres and their stability rather than the multiplex gain.

The use of optical fibres in astronomy is subject to the loss of Etendue known as Focal Ratio Degradation (FRD). This is believed to be caused by small-scale imperfections in the fibre which scatter power between modes. Gloge's (1972) model reproduced the spreading of the output beam but also predicted a dependence on length which was much stronger than that seen in fibre systems used in astronomy. This was explained by Poppett and Allington-Smith (2010) by assuming that a small amount of scattering took place within the bulk of the fibre, but a much greater effect was generated within a small region of fixed length at the ends of the fibre. Thus the dependence of FRD on the fibre length is much diluted. However this begs the question of the physical processes operating in the FRD-generating zone (FGZ). It is not clear if the defects are linked to the relief of the core/cladding boundary or to stress-induced variations in refractive index caused by the fibre termination or a combination of these effects. One critical piece of information is the size of the FGZ. Correlating this information with the measured FRD performance would thereby help to improve the performance of fibres.

We present the results of tests of fibre FRD and try to interpret them in terms of a ray-tracing model previously developed by Allington-Smith, Murray and Lemke (2012; hereafter AML). The failure of this model to reproduce the observed increase in FRD at small injection angles leads us to develop an extension to Gloge's model which fits the data well. Using this we conclude that the length of the FGZ is the critical parameter and that this is very small (similar to the core diameter) for fibres which show low levels of FRD. Finally, we link this to differences in the fibre termination methods used.

# 2 FRD measurements

One of the most useful methods for measuring FRD is to examine the thickness of the ring produced in the farfield when the fibre is illuminated by a collimated beam with a specified (injection) angle with respect to the optical axis of the fibre (Figure 1). In the absence of FRD, rays are scattered only in azimuth and so exit from the fibre with the same radial angle as at the input to form a thin ring in the farfield. When FRD is present, rays are scattered to higher and lower angles with the result that the output ring has a finite width. From this, the effect of FRD can be predicted in a more realistic situation by integrating over the injection angles present when illuminated by a telescope with an appropriate central obstruction.

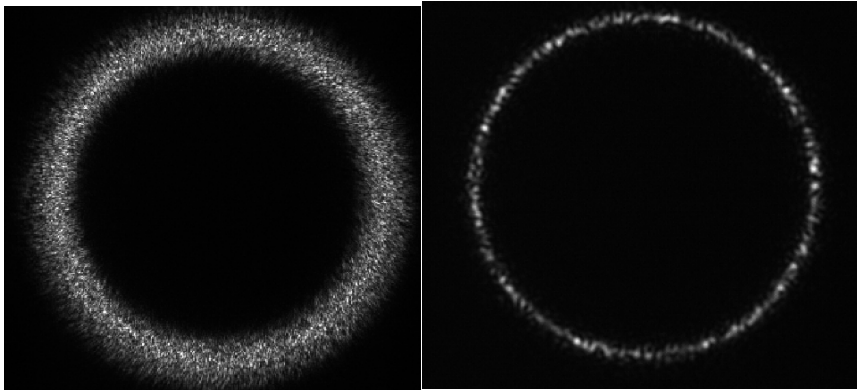

Figure 1: Examples of rings produced with collimated input light. Both are for the same injection angle (10 deg) and are recorded with the same angular scale. The difference between them is due to the different FRD which is much better for the fibre on the right..

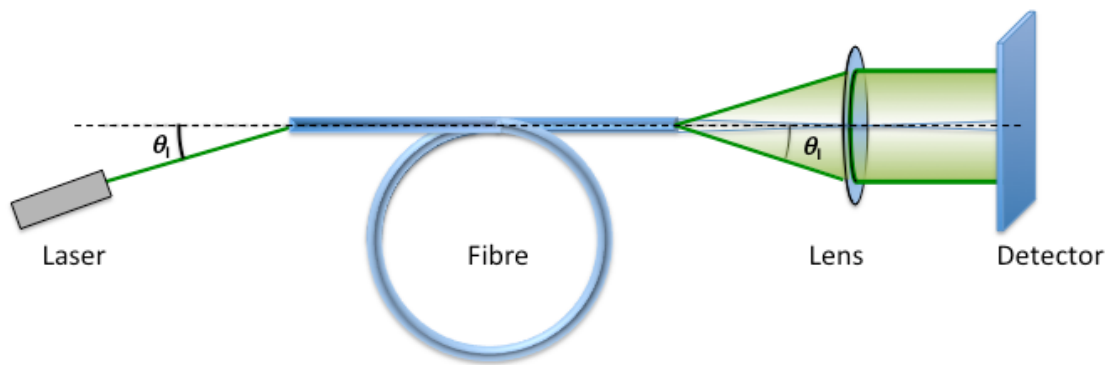

Figure 2: Schematic diagram of apparatus used to measure FRD

The experimental procedure is detailed in AML and indicated schematically in Figure 2. Light from a fixed laser enters the fibre which is mounted in a stress-free chuck attached to a goniometer. Back illumination through the fibre is used to establish the optical axis of the fibre. Both positive and negative angles are used to account for any misalignment between the optical axis and the normal to the polished end. The other end of the fibre is mounted in a similar chuck. Light emerging in a ring is reimaged on the detector (a CCD).

Exposures without illumination are taken to allow background light to be subtracted. Residual background in the illuminated images was removed by subtracting the signal estimated from the unilluminated part of each image. The geometric centre of each ring was determined and the ring profile obtained by summing in azimuth about this point. The error in the determination of the centre is less than 1 pixel and so makes a negligible contribution to the uncertainty in the ring thickness.

One key aspect of the system is that the farfield illumination is recorded directly on the detector instead of by projection onto a screen in contrast to other workers (e.g. Haynes et al. 2008, Eigenbrot et al 2012). As noted by Eigenbrot et al., the use of a diffusing screen can broaden the rings by scattering within the depth of the diffuser.

The fibres were mounted in ferrules using a commercial adhesive and the composite assembly polished using a commercial polishing machine following a standardised process developed by us and the vendor. This generally produces good quality fibres selected with the aid of quality control procedures involving inspection of the fibre ends with a microscope. The data presented below are for the same batch of polymicro FBP fibre with core/cladding/buffer diameter of 120/170/190µm but with different terminations as described below.

## 2.1 DU-1 data

The DU-1 fibre was terminated with a steel ferrule into which it was glued without stripping the buffer. The glue was allowed to penetrate (wick) into the gap between fibre buffer for several mm before curing.

Figure 3 shows details of fits to the DU-1 data using the Gloge model in addition to a simple model of stray light based on a Lorentzian profile centred on the optical axis (AML). The rings were individually summed in azimuth following determination of the geometric centre.

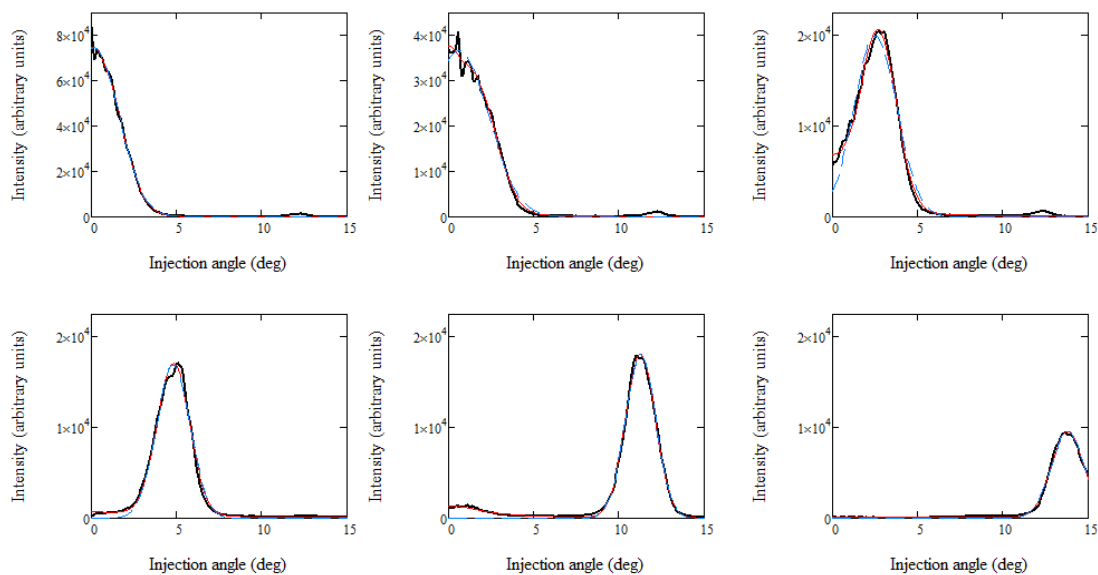

Figure 3: Examples of fits to DU-1 ring profiles data (black) using the Gloge model with an additional Lorentzian scattering component (thin red line) and with a Gaussian function (dashed blue line). A selection of rings are shown for injection angles, 1, 2, 3, 4, 11 and 14 deg. In most cases, the curves overlay each other so closely that the individual traces cannot be distinguished.

The model parameters in the fit are the peak intensity, the diameter of the ring, the FWHM of the ring, the relative amount of scattered light and the width of that distribution.

Gambling et al. (1975) showed that Gloge's model produced a single Gaussian function for on-axis rays (zero injection angle) and a ring with a Gaussian profile at larger injection angles. At intermediate angles (2-4 deg; depending on the ring thickness) the distribution resembles a broadened single Gaussian function which spans the geometric centre, presenting a problem for fitting. However our fits with the Gloge model are very good even in these cases.

Although, at large injection angles (up to the Limiting Numerical Aperture – see AML) the FWHM of the best-fitting Gaussian function is a useful empirical measure of FRD, it is better to fit the ring thickness using the Gloge model to find the $d_0$ parameter which describes the number-density of scatters (AML).

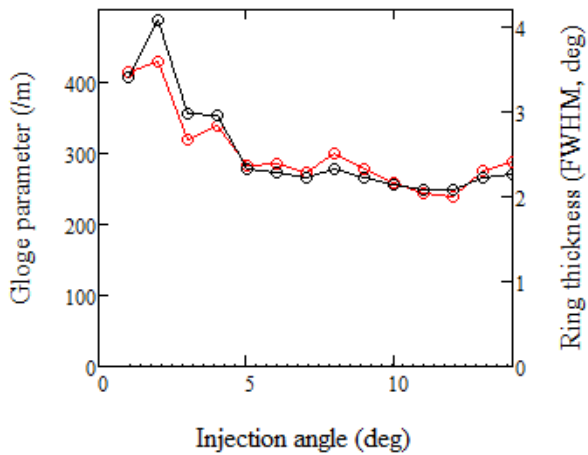

Figure 4: Fitted $d_o$ parameter to DU-1 data (black; left scale) compared to FWHM determined from Gaussian fit (red) right scale).

Examples of fitted data are shown in Figure 3 for various injection angles. These show that stray light is present in some of the data, mostly concentrated on the optical axis at levels of 0-10%, but the ring profiles themselves show very little evidence for scattered light. The profiles are nearly Gaussian with little evidence for extended wings, indicating negligible scattering from the fibre end-face. This finding differs from that of Haynes et al who find it necessary to introduce a component of the PSF with a Lorentzian profile to account for scattering. Eigenbrot et al. also report such a component although this affects only fibres polished with coarse grits implying that surface scattering is not a significant contribution to FRD in normal circumstances.

We also fitted Gaussian profiles for consistency with the other datasets. The difference between the fitted $d_0$ parameter and the FWHM is negligible (Figure 4). In particular, the decrease in ring thickness with increasing injection angle (see below) is not caused by the details of the fitting function. The uncertainty in the ring FWHM was estimated as 0.15 deg.

## 2.2 DU-2 data

The DU-2 fibres were terminated with a ceramic ferrule designed to allow the length of the adhesive bond along the fibre to be minimised to ~1mm compared to several mm for the DU-1 fibres.. Data were obtained by averaging the ring thickness of 17 fibres illuminated with collimated beams injected at angles 5-14 deg to the input optical axis. Each fibre was measured with light injected at both ends and with the input beam at positive and negative angles to account for any misalignment in angle between the core and the fibre mount (making 58 data points in all). The uncertainty in the ring thickness was obtained from the standard error assuming 17 independent measurements. Figure 5 shows the derived ring thickness for both datasets. It can be seen that there is a very large difference in FRD. The DU-2 results are exceptionally good while the DU-1 fibres are quite typical for fibre systems to date.

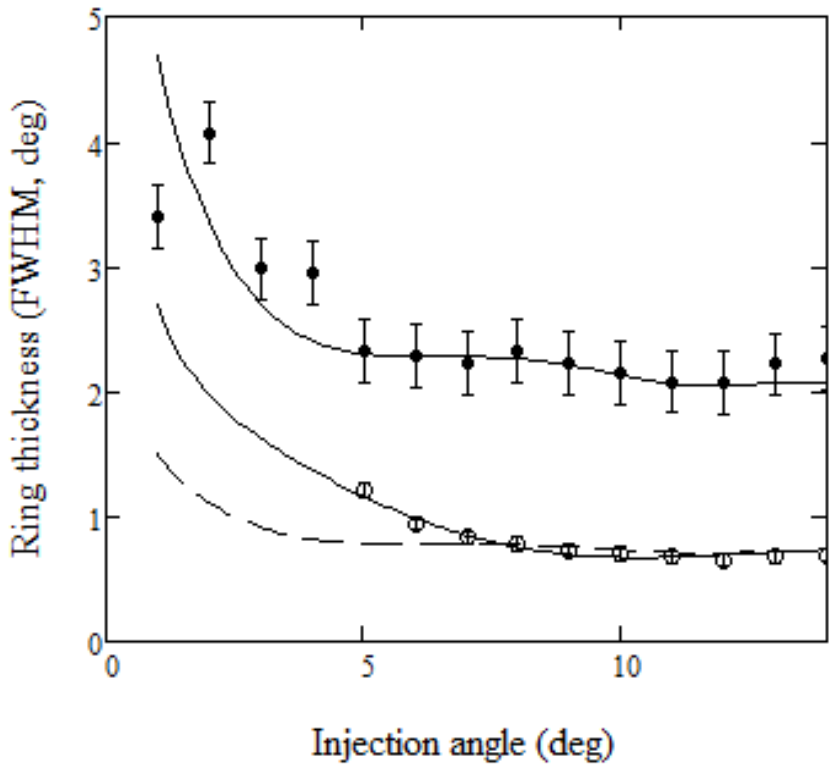

Figure 5: Ring thickness for the datasets described in the text: filled circles: DU-1 fibre; open circles: the mean of 17 DU-2 fibres (error bars smaller than the symbol size). The curves are fits described in the text.

The striking feature in both datasets is the increase in ring thickness with decreasing injection angle, most evident at <5 deg. This is present in both sets of data despite the large difference in ring thickness between them. It should be noted that the ring thickness at small injection angle does not vary smoothly with injection angle. The excellence of the individual fits to the Gloge model (Figure 3) implies that is not a problem with the fitting process. This phenomenon is also noted by Eigenbrot et al. (2012).

# 3 Model of FRD generation

## 3.1 Ray-tracing model

To help understand these and other questions about optical fibres in astronomy, we use the *FibRay* model based on ray-tracing (AML). The geometry of the ray-tracing is shown in Figure 6. This exploits the fact that fibres of interest mostly operate in the extreme multimode limit; when the number of modes present in the fibre is

$$M \approx 2\pi^2 \left(\frac{R\Theta}{\lambda}\right)^2 \quad (1)$$

(e.g. Snyder and Love 1983) where $\Theta$ is the Limiting Numerical Aperture (LNA) determined by the contrast in refractive index between the core and cladding, $R$ is the core radius and $\lambda$ is the wavelength. For $2R = 120\,\mu m$, $\Theta = 0.22$, $\lambda = 1\mu m$, the number of modes is $M \sim 3500$ which implies that one should be able to simulate the behaviour of a real fibre by tracing a few thousand of rays.

FRD is simulated by perturbing the angle of reflection of rays that strike the core/cladding boundary within the FGZ which has dimensionless length, $z_F = L_F/R$. The perturbation follows a Gaussian function which is independent in the $x$ and $y$ direction but has the same characteristic size, $q_F$. As reported in AML, this results in FRD which is very close in form to that observed and can be made to reproduce the observation by a one-time calibration.

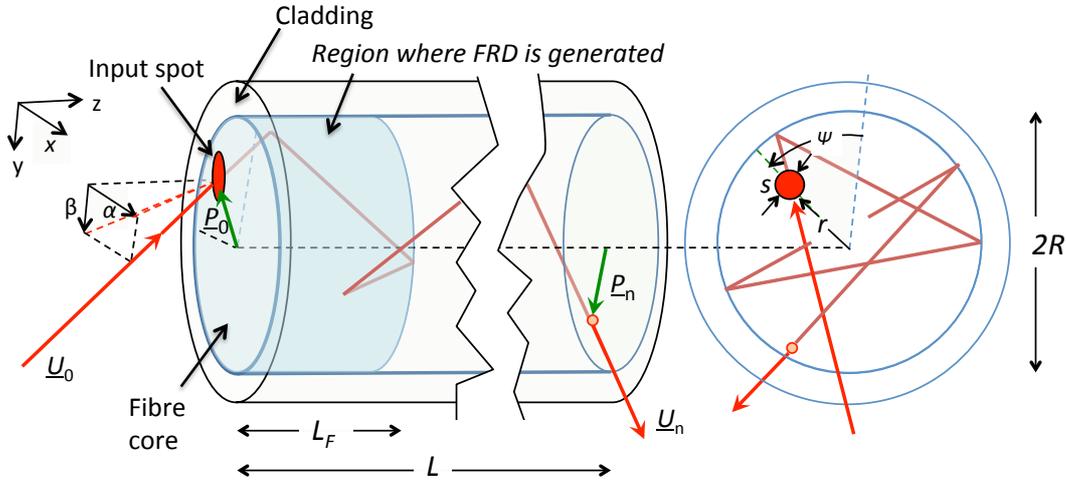

Figure 6: Geometry of FibRay simulation. The sketch shows how a light ray enter the fibre along vector **U₀** via a defined spot and propagates along the fibre until it exits along vector **Uₙ** after $n$ reflections. In this case, the rays are skew since they are not restricted to a meridian such as that defined by the position vector **P₀** which represents the initial intersection of the ray with the fibre input face. The shaded region indicates the FGZ.

An example of output from *FibRay* in this configuration is shown in Figure 7. This shows the ring in the farfield while the nearfield output distribution is quite flat.

The *FibRay* model initially used $z_F = 5$ for each end of the fibre and $q_F = 0.014$. This was chosen because it produced a ring thickness of 1.4 deg FWHM typical of the fibre with diameter $2R = 120\mu m$ which we had tested.

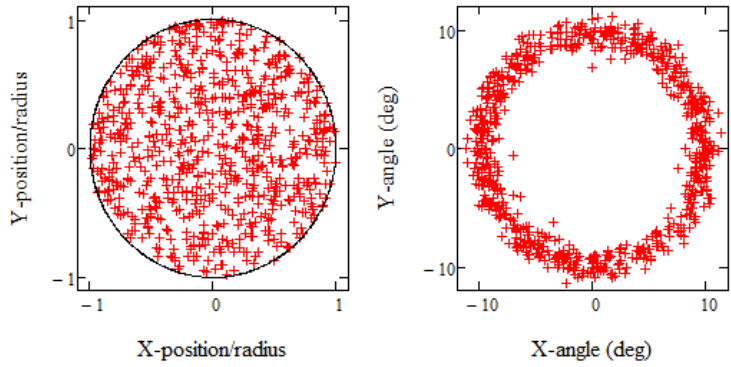

Figure 7: Spot diagrams for nearfield (left) and farfield (right) from the FibRay simulator for illumination by a collimated beam injected at angle 10 deg to the fibre axis and filling the input face of the fibre. For clarity, only 1000 rays have been traced.

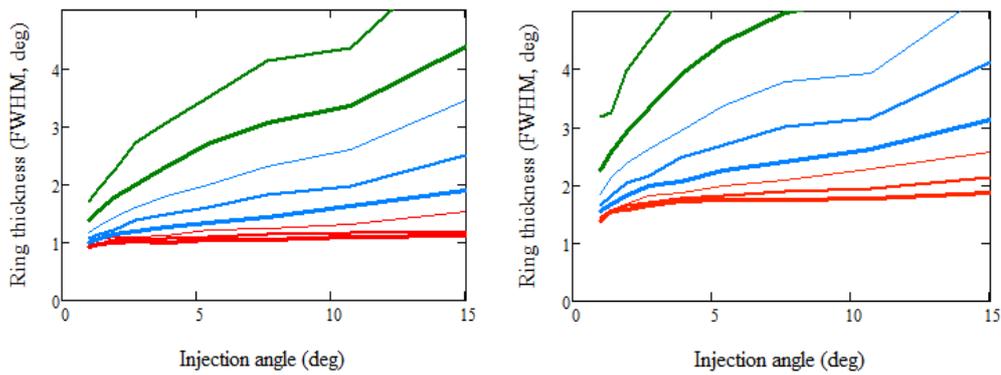

Figure 8: Ring thickness predicted by the FibRay model versus injection angle for different values *of* $z_F$ increasing logarithmically from $z_F$ = 0.5 (thick red) to 100 (medium green). Left – for $q_F$ = 0.015, right – for $q_F$= 0.025

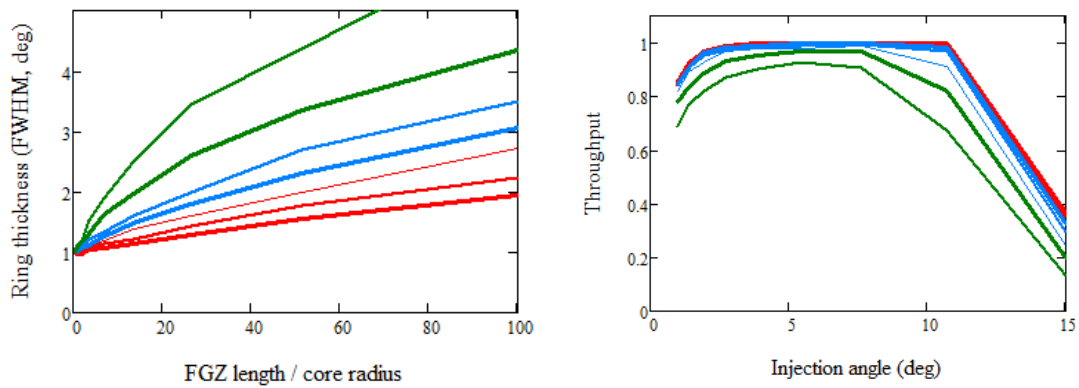

Figure 9: Left – Predicted ring thickness versus FGZ length for for $q_F$=0.015 for different injection angles increasing logarithmically from 1 (thick red) to 15 deg (medium green). Right – Predicted throughput versus injection angle for the same case.

However it was noted that the ring thickness varied when other values of $z_F$ were used. Figure 8 shows the predicted ring thickness, $w$, as a function of injection angle (which defines the ring radius) for different value of $z_F$. It can be seen that *w increases* with injection angle, opposite to the observed behaviour, in all cases where $z_F > 2$. This is not true for smaller FGZ lengths which is why a small FGZ length was initially chosen.

The effect of varying $q_F$ can also be seen in Figure 8. The near-degeneracy between $z_F$ and $q_F$ means that it is possible to keep ring thickness constant for higher $z_F$, but only by changing the values of $q_F$ to suit, This is unphysical unless one assumes that the angular scattering profile varies with angle.

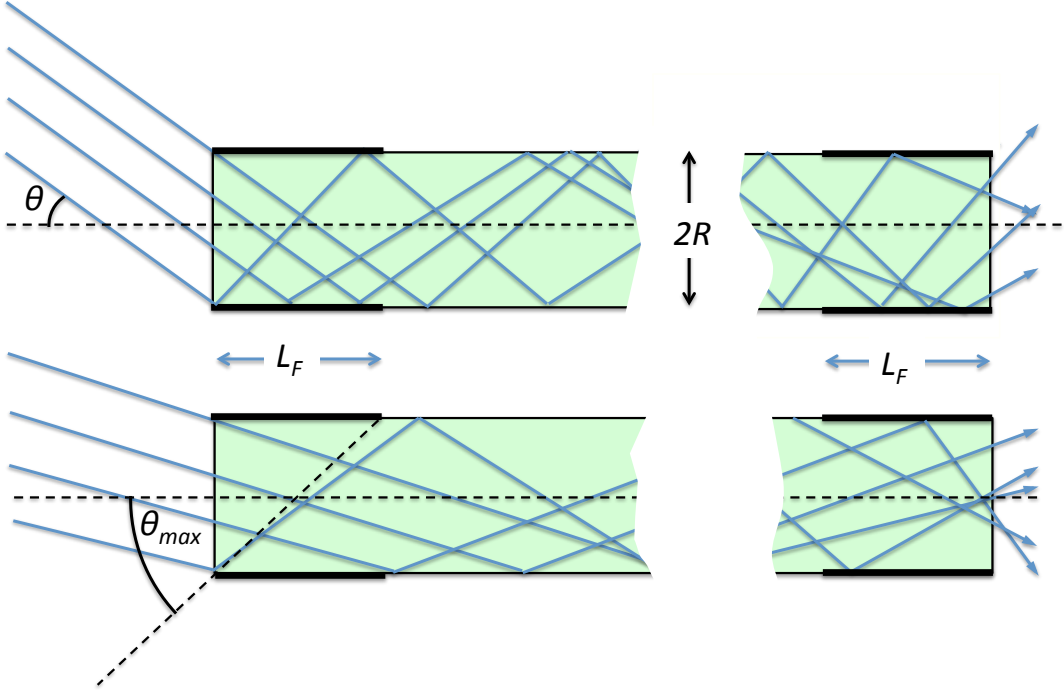

Figure 10: Scattering geometry for meridional rays with different injection angles. The core/cladding boundary in the FGZ are shown as heavy black.

To understand these results better, consider the geometry in Figure 10. This shows that light entering the fibre will strike the FGZ only if it has a steep angle or if it strikes near the edge of the input face. Rays with small radial angles are likely to avoid the perturbing effect of the FGZ . We can define a critical FRD angle (not to be confused with the critical angle which defines the LNA) at which half of the rays will be scattered using the expectation value of the injection angle

$$\theta_F = \frac{\int_0^{\theta_{max}} \theta p(\theta) d\theta}{\int_0^{\theta_{max}} p(\theta) d\theta} = \frac{2}{3}\theta_{max} \approx \frac{4}{3z_F} \qquad (2)$$

since the probability distribution $p(\theta) \propto \theta$ due to the azimuthal symmetry about the optical axis and $\theta_{max} = 2/z_F$ and we have used the small angle approximation. The same restriction affects rays leaving the fibre at the other end. The number of FRD-generating bounces is therefore

$$N = \frac{z_F}{z} = \frac{3}{4}z_F\theta \qquad (3)$$

Note that this analysis does not include skew rays which may experience more bounces. Examination of the data suggests that the locus of datapoints corresponding to the critical angle is approximately where the ring thickness is 2 deg. Thus on all occasions when the ring thickness, $w > 2$deg, each ray will experience at least one bounce. Below this thickness, the number of bounces within the FGZ is restricted

In the model, the probability of obtaining a given perturbation is

$$p(\theta_j, \theta'_j) \propto \exp\left[-2\frac{(\theta_j - \theta'_j)^2}{q_F^2}\right] \qquad (4)$$

where $j = x$ or $y$ to indicate the angle measured from the local $x$ and $y$ axes and $q_F$ is a measure of the variation in the perturbing angle. After $N$ such bounces the angular distribution may be expected to be broadened by a factor $\sqrt{N}$ so we expect the ring thickness

$$w \propto \sqrt{z_F \theta} \qquad (5)$$

In Figure 9 we see that the ring thickness does indeed scale approximately with $\sqrt{\theta}$ (for large $w$) but this is only valid above $w > 2$ deg. For smaller injection angles, the number of bounces remains constant at $N = 1$, so no increase in ring thickness is seen. Furthermore the ring thickness scales roughly with $\sqrt{z_F}$ for fixed injection angle as predicted.

This behaviour is not caused by mode-stripping, which is analogous to a large-scale loss of rays which have been scattered to angles exceeding the LNA and so are lost into the cladding. Figure 9 (right) shows the predicted throughput against injection angle. Except for rays *injected* at angles exceeding the LNA (which corresponds to 12 deg), but not necessarily subsequently scattered to large angles, light is lost only for very large values of $z_F$ where some fraction of scattered rays will exceed the critical angle.

While this is a satisfactory explanation for the absence of a rise in ring width with increasing injection angle, it does not explain the increase at small injection angles. For this we need to reconsider wave-based models since ray-tracing cannot correctly account for the modal properties of fibres in all cases.

## 3.2  Wave model

FRD was modelled by Gloge (1972) as a modal diffusion process whereby energy can be scattered to the nearest waveguide modes. These modes are analogues of sets of rays propagating along the fibre with similar angles. In general their direction vectors have a radial and azimuthal component to match the radial and azimuthal indices of the waveguide eigenmodes.

The full analysis requires the formulation of a diffusion equation (Snyder and Love 1983) which can be solved in terms of Laguerre polynomials if certain simplifications are made. These include the assumption that the parameter describing the scattering has no angular dependence and that the range of integration can be extended from (0, $\theta_C$) to (-∞,∞) where $\theta_C$ is the critical angle defined by the LNA. Also it assumes that the scatterers are uniformly distributed throughout the medium. As we have seen, there is evidence that the scatterers are highly concentrated at the ends.

We can understand how this inhomogeneity may affect the solution using insight obtained from ray-tracing. We can generalise the FRD critical angle defined in equation (2), by noting that the number of bounces of meridional rays is $N = z_F/z$ where $z$ is the distance travelled parallel to the optical axis between bounces. Thus the critical FRD angle for $N$ bounces is

$$\theta_F(N) = \frac{4N}{3z_F} \tag{6}$$

so that meridional rays propagating at angle $\theta$ will experience between $N$ and $N+1$ bounces inside the FPZ if $\theta_F(N) < \theta < \theta_F(N+1)$.

The number of rays incident at angle $\theta$ scattered into angle $\theta'$ (*Figure 11*) is then

$$p(\theta',\theta) \propto (\theta'-\theta)\exp\left[-\frac{2(\theta'-\theta)^2}{q_F^2}\right] \tag{7}$$

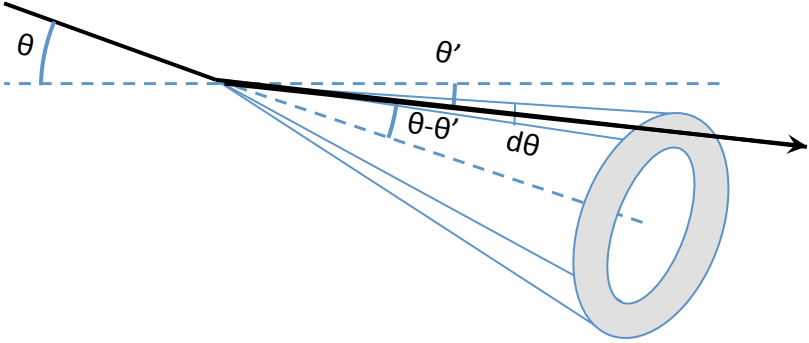

*Figure 11: Geometry of a scattering event*

After $N$ scatterings, the initial distribution of angles around the initial injection angle, $\theta = \theta_I$, will be be broadened by $\sqrt{N}$. The expectation value of $\sqrt{N}$ is

$$\langle\sqrt{N}\rangle \approx \frac{\int_0^\infty p(\theta',\theta)\sqrt{N}\,d\theta}{\int_0^\infty p(\theta',\theta)\,d\theta} \approx \sqrt{\frac{3}{4}z_F}\frac{\int_0^{\theta_C}\sqrt{\theta}(\theta-\theta')\exp\left[-\frac{2(\theta-\theta')^2}{q_F^2}\right]d\theta}{\int_0^{\theta_C}(\theta-\theta')\exp\left[-\frac{2(\theta-\theta')^2}{q_F^2}\right]d\theta} \tag{8}$$

using the number of bounces in the FGZ from equation (6) on the assumption that the scattering is small, $|\theta - \theta'|<<\theta$, and that the upper integration limit is the critical LNA angle.

Although we can estimate the broadening factor, we still need to know the initial thickness of the ring before scattering. Using the basic premise of the Gloge model, that power is only transferred between adjacent modes (which he supports with experimental data), we can estimate this using the wave model. He further assumes that there is a continuum of modes throughout the fibre but this assumption can be expected to break down at small propagation angles where the fibre may behave like a single-mode or few-mode fibre. This causes an uncertainty in the injection angle

$$\Delta\theta_I = \left(\frac{d\theta}{dM}\right)\Delta M = \frac{1}{\theta_I}\left(\frac{\pi\lambda}{R}\right)^2 \qquad (9)$$

where $\theta_I$ has been substituted for the LNA in equation (1) and the uncertainty in the mode number is set to unity. The uncertainty in the injection angle can be translated into ring thickness via $w_0 = \sqrt{\Delta\theta_I}$ since only the radial direction is affected. Thus the final ring thickness is $w = \langle\sqrt{N}\rangle w_0$ where $\langle\sqrt{N}\rangle$ is given by (8).

In summary, the model has two components: the broadening of the radial PSF (i.e. the thickness of the rings) due to repeated scattering events whose number increases with both the length of the FRD-generating zone and the injection angle; and the discreteness of the waveguide modes at small injection angles.

Fits to the DU-1 and DU-2 data are shown in Figure 5. The model (solid lines) fits the data well except at the very smallest injection angles where mode discretisation may be responsible for some of the scatter.. The approximate effect of diffraction has been included by adding in quadrature a Gaussian function with FWHM $\lambda/2R\cos\theta_I$ where the dependency on $\theta_I$ accounts for the reduction in effective aperture for non-zero incident angle. This has very little effect in practice since its value is typically a few-tenths of a degree.

*Figure 12*, shows contours of the logarithm of the reduced $\chi^2$ as a function of $z_F$ and $q_F$ parameter. The quality of fit may be judged by eye from Figure 5 but the minimum of the reduced $\chi^2$ is 3.2 for the DU-1 data and 0.7 for DU-2. The difference between the DU-1 and DU-2 results is due mainly to a large increase in the length of the FGZ confirming the interpretation obtained from ray-tracing. The best fitting $z_F$ values are 24±4 and 140±16 for DU-1 and DU-2 respectively.

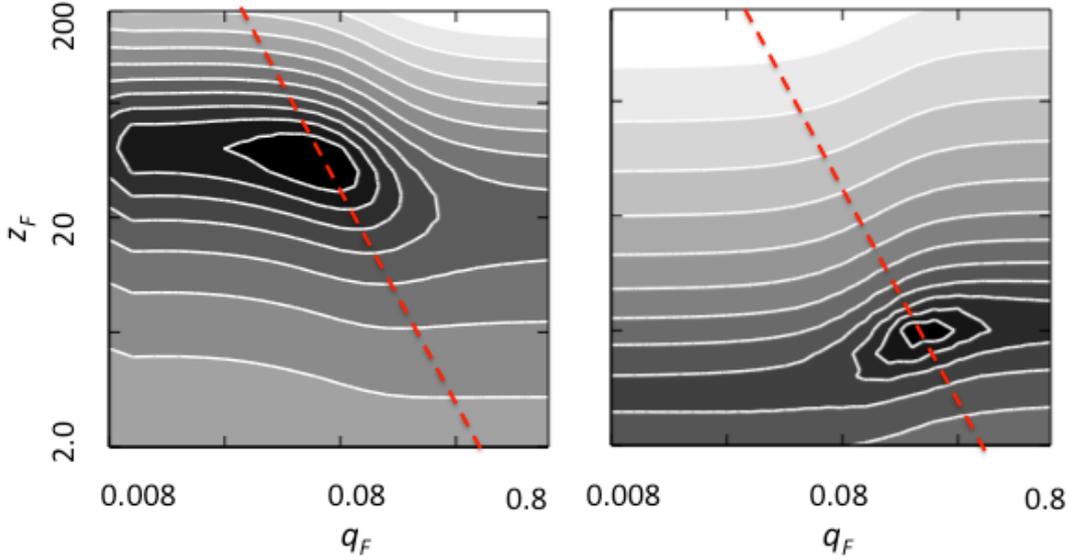

*Figure 12: Contours of the logarithm of the reduced $\chi^2$ for the fits described in the text for the DU-1 (left) and DU-2 (right) respectively. The optimum values correspond to the minimum which is shaded black. The red dashed line indicates $q_F^2 z_F$ = constant.*

The best fitting values for $q_F$, which controls the detail of the upturn at low injection angle are 0.16±0.02 for DU-1 and 0.07±0.02 for DU-2 respectively. The fits suggest a relationship $q_F \propto 1/\sqrt{z_F}$. The small inferred FGZ length for the DU-2 fibres implies that the scattering is confined to a length ~1.5mm. For the DU-1 fibres, the length is ~10mm.

## 4 Discussion

### 4.1 Implications for fibre preparation

Our results imply that for fibres with good FRD, the FGZ is very short, ~1.5mm, but much longer, ~10mm, for fibres with poor FRD. We see no evidence for surface scattering in agreement with Eigenbrot et al. who see it only in fibres finished with a coarse grit, but in disagreement with Haynes et al.

There are two main physical mechanisms that may be proposed to explain FRD.

- Stress induced in the fibre core/cladding boundary by the termination process. This could cause small-scale physical deviations. However since these should be present over the entire length of the fibre, it is necessary to assume that stress (either uniform or variable on small scales) can turn these physical inhomogeneities into variations in refractive index which could scatter the light or perturb the ray trajectories.

- Scattering due to the physical roughness of the end-faces. This would perturb the trajectory of rays passing through the end-faces. In addition, the effect of stress could further perturb the rays by changing the refractive index.

For the first mechanism, we may expect it to depend on the detail of the fibre termination. This involves insertion of the fibre into a cylindrical sleeve (ferrule) of length 1-10mm and securing by small amounts of adhesive which penetrate (by capillary motion) from the fibre end along the length of the sleeve. The adhesive is then cured (usually by heating since ultraviolet light cannot penetrate into the ferrule). Many different adhesives are available, some of which are claimed to be dimensionally stable when heated – which might be expected to exert less stress on the encapsulated fibres.

Surface roughness and stress may also be produced in the polishing process. Once the fibre is encapsulated in the ferrule, its end face is polished using successively finer grits in a slurry with a minimum particle size of ~0.1µm. Thus we may expect the size scale of imperfections to be on this scale although variations in stress may affect larger regions.

Scattering from the end-face, if present – see above – may be reduced by the common practice of immersing the fibre end to a glass plate. With an index-matching medium between the fibre and the plate, the effect of physical perturbations in the fibre end-case should be eliminated or much reduced.

The concentration of FRD-generation at the fibre ends where stress is most likely to be caused and frozen-in by the encapsulation process, supports the idea that stress is the most important effect, perhaps acting in concert with microscopic defects in the core/cladding boundary relief. The encapsulation process necessarily involved the use of adhesives which may be dimensionally unstable during the curing process thereby causing stress.

Our results suggest that the primary mechanism is related to the length of the FGZ, $z_F$, instead of the width of the scattering function, $q_F$. It is interesting that the inferred length of the FGZ is similar to the differing lengths of the bond actually used in the DU-1 and DU-2 fibres.

This view contrasts with that of Eigenbrot et al. who demonstrate a strong link between the grit size used in the polishing and FRD, implying that it is the width of the scattering function that is the dominant factor. However this is not necessarily in disagreement with our model. Our normal polishing procedure tends to produce a low-level of bulk scattering in the FGZ, with the result that our results are now more sensitive to the length of the FGZ. Furthermore, as already noted, the FRD parameters are partly degenerate and our conclusion are based on arguments related to the detailed quality of the model fit. Our conclusions could also be inverted by arguing that the greater sensitivity of FRD to changes in $q_F$ argues in favour of the scattering width being the dominant parameter.

Although few would disagree that the strength of the scattering, i.e. the amplitude of the scattering angle, as parameterised by $q_F$, plays a major role in determining FRD, our results, taken together with that of Poppett and Allington-Smith (2010) provides strong support that the length of the FGZ also plays a pivotal role by determining the number of scattering events.

As noted in §2, for the DU-2 fibres, great care was taken to reduce the length of fibre in contact with the fixing adhesive. This may indeed have resulted in the minimal contact length which our analysis implies. If so, it is clear that the secret to reducing FRD is to reduce the amount of adhesive used in the encapsulation.

## 4.2  Implications for fibre models

Explaining the complexity of FRD phenomena exposes limitations in the *FibRay* model which is based on ray-tracing despite its excellence at simulating most features of fibres used in astronomy. Although these fibres operate in the multimode-limit where there is a continuum of modes, this approximation appears to break down for the small regions where FRD is generated. The wave model of Gloge fares no better in these demanding circumstances but we have been able to extend the model so that it provides a satisfactory fit to the data.

Note that both the model fit is not dominated by the mode discretisation. This is demonstrated by the difference between the best-fit to the DU-2 data and the dashed curve in Figure 5 which corresponds to $z_F$ = 24 and *$q_F$ = 0.07*, i.e. the value of $z_F$ that best fits the DU-2 data at large injection angle but with the smaller value of $q_F$ that best fits the DU-1 data.  The shape of the curves not only differs from the $1/\sqrt{\theta_I}$ relationship expected from mode discretisation but the difference in shape suggests that there is not a single mechanism at work.

## 5  Conclusions

By studying the variation of FRD with injection angle, we have developed a new model to predict the thickness of rings produced in the farfield with collimated illumination and explained other puzzling phenomena encountered in fibre testing. These data are commonly used to assess the quality of fibres used in astronomy where limiting performance is critical for upcoming cosmology projects such as DESI.

The new model suggests that the observed variation of ring thickness with injection angle is consistent with variation in the small region at the ends of the fibre where they are terminated. For fibres with good FRD, this region appears to be very short, 1-2mm, but ~10mm for fibres with poorer FRD. This picture is consistent with the difference in fibre preparation in which the good-FRD fibres were encapsulated in ferrules of an advanced design that allowed the amount of adhesive into which the fibre came into contact to be reduced to very low levels. Indeed, the inferred length of the FRD-generating region is consistent with the actual length of the bonded region.  We conclude that the key to improving FRD is to reduce the amount of adhesive.

## Acknowledgements

We thank our colleagues at the Lawrence-Berkeley National Laboratory at the University of California working with us on the Dark Energy Spectroscopic Instrument (DESI): Jerry Edelstein and Claire Poppett, for helpful discussions. We also thank Will Saunders for  helpful  and illuminating discussions regarding the theory of FRD.  We acknowledge support from the UK Science and Technology Facilities Council.## References

Allington-Smith, J.R., Murray, G.M. and  Lemke, U. 2012. MNRAS, 27, 919-933

Colless, M. et al.  2001. MNRAS  328, 1039.

Eigenbrot, A.D, Bershady, M.A. and Wood, C.M. 2012. ArXiv: 1208.0829v1

Gambling. W.A., Payne, D.N. and Matsumura, H., 1975. Appl. Opt., 14, 1538.

Gloge, D. 1971. Appl. Opt 10, 2252-2258.

Haynes, D.M., Withford, M.J., Dawes, J.M., Haynes, R and Bland-Hawthorn J,B., 2008. Proc. SPIE 7018, 70182U.

Kimura, M. et al. 2010. PASJ. 62, 1135.

Poppett, C. and Allington-Smith J., 2010 MNRAS, 404, 1349

Poppett, C.L 2011. PhD thesis, Durham University.

Schlegel, D. et al. 2011. arXiv e-print: 2011arXiv1106.1706S

Snyder, A.W. and Love, J.D 1983. "Optical waveguide Theory", Kluwer.